# Mode-locked fiber laser with offset splicing between two multimode fibers as a saturable absorber

Zhipeng Dong, Shujie Li, Peijun Yao, Chun Gu, Lixin Xu*

*Department of Optics and Optical Engineering, University of Science and Technology of China, Hefei 230026, China*

*Corresponding author: xulixin@ustc.edu.cn

**Abstract**: A novel mode-locked fiber laser based on offset splicing technology between two kinds of graded index multimode fibers is proposed and experimentally demonstrated. The offset splice spot structure acts as a saturable absorber via nonlinear multimodal interference mechanism. The absorption modulation depth of the device is 2.8%, with a saturation intensity of ~2.5 MW/cm$^2$. The fiber laser operates at a stable mode-locked state when the pump power reaches 114 mW. The laser oscillates at 1039 nm with a pulsewidth of 17.6 ps, a repetition rate of 32.619 MHz, and a high signal noise ratio of ~65 dB. The fiber laser is superior in terms of low cost, high damage threshold, and easy fabrication, thus providing many potential applications in ultrafast photonics and industrial processing.

## 1. Introduction

Ultrafast laser takes act an important role in numerous fields of modern technology and is an indispensable research tool for optical frequency combing [1], nonlinear optics [2-3], and material processing [4]. Mode-locking is a main method to obtain ultrafast laser sources. In theory, to obtain a stable mode-locked laser, the so-called saturable absorption property of materials through which the light absorption decreases with the increase of light intensity plays a decisive role [5]. A number of saturable absorption materials, i.e., saturable absorbers (SA), have been widely applied in fiber lasers, such as $Bi_2Te_3$ [6], black phosphorous [7-9], graphene [10], $WS_2$ [11] and so on. However, these materials often suffer from either relatively low damage thresholds or high manufacture cost, which hinders their practical applications.

Compared with traditional step index multimode fiber, gradient index multimode fiber (GIMF) has a lower mode dispersion, less bending loss and wider range of operation, which are suitable for short distance communication networks and high-speed data transmission. Recently, GIMF has been extensively studied because of its unique nonlinear multimode interference (NLMMI) characteristics. New nonlinear dynamics in GIMF such as supercontinum generation [12-14], spatiotemporal mode-locking [15-16], soliton molecules [18], self-beam cleaning [18], and second harmonic generation [19-20], have been reported recently. As early as 1994, the GIMF has been used as a bass-band filter in all-fiber lasers [21]. In 2003, E. Nazemosadat and A. Mafi have theoretically suggested that the single mode fiber (SMF)-GIMF-SMF structure could act as a SA in all-fiber mode-locked lasers [22]. The SA based on NLMMI property has a clear superiority of simple structure and easy fabrication as well as high damage threshold (the damage threshold of quartz), especially suitable for high-energy pulse fiber mode-locked lasers. However, according to the theory proposed by A. Mafi [22], the length of the GIMF in the SMF-GIMF-SMF structure needs to be precisely controlled in tens of microns scale, which is hard to be realized in practice. Lately, Z. Wang et al. propsed a SMF-SIMF-GIMF-SMF structure as SA, which has also been experimentally verified in Tm-doped and Er-doped all-fiber soliton laser [23]. By adopting a short section of SIMF, one can eliminate the restriction on the GIMF length. In addition, F. Yang *et al.* also experimentally demonstrated the feasibility of using SMF-GIMF-SMF structure with an inner micro-cavity as a SA [24].

Here, we present a SMF-GIMF-GIMF-SMF structure with offset splice spot between two kinds of GIMFs with different core-diameteras a SA based on the multimode nonlinear interference in a multimode fiber. In this SA, a section of ∼ 5cm GIMF (50/125, Corning) was offset spliced with a 23cm GIMF (62.5/125, Corning). Introducing the offset splicing spot can eliminate the restriction on the GIMF length, which is significant in actual fabrication. The absorption modulation depth of the SA is 2.8%, corresponding to a saturation intensity of ∼2.5 MW/cm$^2$. We further use this structure to built an ultra-fast all-laser source with a pulse-duration of 17.6 ps and a repetition rate of 37.619 MHz.

## 2. Fabrication and performance of the SA device

Figures 1(a) and 1(b) show the structural diagram of the device and the image of offset splicing spot by fusion splicer (Fujikura FSM-100P), respectively. When the signal light transmits in the GIMF, various high-order modes will be excited and periodic interference transmits in the GIMF, which is called as the self-image effect. Affecting by the self-phase modulation (SPM) and cross-phase modulation (XPM), the self-focusing length of high-power light is different from that of low-power light. The intensities of XPM and SPM are affected by the following four factors: the number of excited modes $M$ in the GIMF, the mode-field overlapping area ratio between fundamental mode of the SMF and the LG$_{00}$ mode in the GIMF $\eta$, the length of the GIMF $L$, and the total optical power $P$ [22, 25]. The Kerr effect of NLMMI in the GIMF can be described similarly to the description of the self-imaging effectas mentioned in Ref. [26]. When the light couples from the single mode fiber to the GIMF, a specific set of high-order modes of GIMF are excited.

$$E_{SMF}(r,\phi,z=0) = \sum_{m=1}^{M} C_m e_m(r,\phi,z=0) \tag{1}$$

where z = 0 is the splicing spot between the input SMF and GIMF, $E_{SMF}(r,\phi,z=0)$ is the fundamental mode of the SMF and $e_m(r,\phi,z=0)$ is the *n*-th guided mode of GIMF. $C_m$ is the mode expansion coefficient. $M$ is the number of excited modes in the GIMF fiber.

The field $E_{GIMF}(r,\phi,z)$ along the GIMF can be expressed as

$$E_{GIMF}(r,\phi,z) = \sum_{m=1}^{M} C_m e_m(r,\phi,0) e^{-i\beta_m z} = e^{-i\beta_1 z} \sum_{m=1}^{M} C_m e_m(r,\phi,0) e^{-i(\beta_m - \beta_1)z} \tag{2}$$

where $\beta_1$ and $\beta_m$ are the propagation constants of the fundamental mode and the *m*-th excited mode of the MMF, respectively.

If the following condition is satisfied for all $M$ modes, Self-imaging occurs at some certain positions in the GIMF,

$$(\beta_m - \beta_1) Z_{\text{self-imaging}} = \Delta\beta_m Z_{\text{self-imaging}} = n_m 2\pi. \tag{3}$$

Where $n_m$ is an integer and the length of the GIMF segment $L$ is chosen to allow self-imaging to occur at a certain wavelength, Eq. (3) can be rewritten as follow

$$(n_{eff,m} - n_{eff,1}) L = \Delta n_{eff,m} L = n_m \lambda_0. \tag{4}$$

Because of the Kerr effect, the refractive index of the optical fiber depends on the light intensity

and thus Eq. (4) can be expressed as below

$$\Delta n_{eff,m}(I)L = n_m \lambda. \tag{5}$$

According to Eq. (5), the transmission spectrum of the device will change due to the Kerr effect of NLMMI. As a consequence, the transmission of the device at a certain wavelength exhibits a performance similar to a SA. Using offset splicing spot between two different core-diameter GIMF structure can increase significantly the total number of higher order-modes in the GIMF [27-29]. Bending the GIMF can change the mode field distribution and overlapping mode-area of the optical fiber accordingly [23, 24, 30]. When an appropriate bending state of the SA is reaching, the transmittance of the device will be adjusted. The optical signal of the high peak-power could then propagate at low loss from the MMF to the SMF, whereas the optical signal of the low peak-power experiences high loss during the transmittance. Thus, it has the characteristics of a SA.

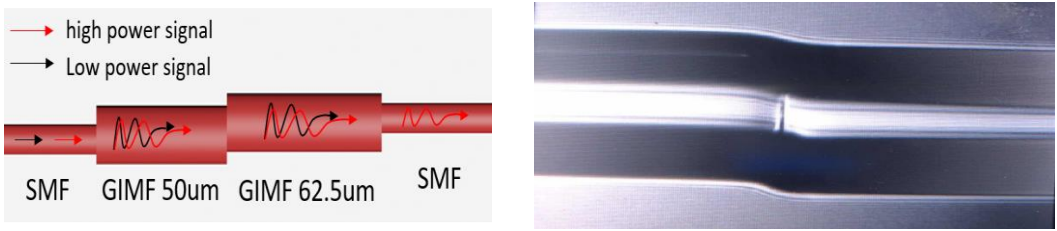

Fig. 1. (a) Schematic diagram of the proposed offset splice spot between different core-diameter GIMF structures; (b) Microscopy image part of the SA

The absorption modulation depth of the SA was measured by a 1.0 um Yb-doped pulse fiber laser. According to the method described by J. Du et al. [14], the transmission curve can be fitted:

$$T(I) = 1 - \alpha \times \exp(-I/I_{sat}) - \alpha_{ns}. \tag{6}$$

where $T$ is the transmittance, $\alpha$ is the modulation depth, $I$ is the input light intensity, $I_{sat}$ is the saturation intensity, and $\alpha_{ns}$ is the nonsaturable loss. As shown in Fig. 2, the absorption modulation depth of 2.8%, corresponding to a saturation intensity of ∼2.5 MW/cm$^2$, which is comparable to the other SAs proposed previously [6-11]. The low saturable intensity characters may have potential applications in ultrafast optics and mode-locked fiber laser.

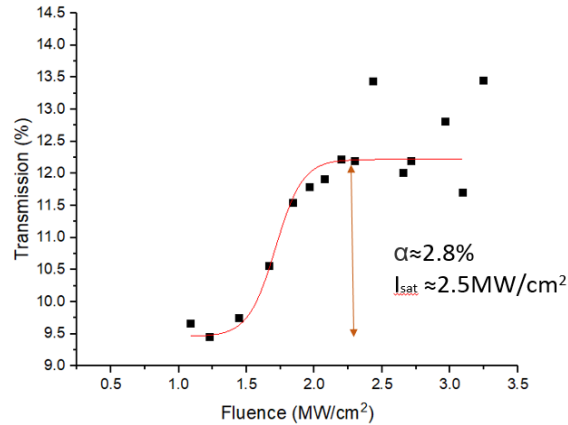

Fig. 2. The measured data of transmission and its fitting curve.

3. **Experimental setup and performance evaluation of the fiber laser**

The schematic of the fiber ring laser is illustrated in Fig. 3. The fiber laser was pumped by a 976 nm laser diode with a maximum power of 450 mW. A section heavily Yb-doped fiber (LIEKKI YB1200, with 1200dB/m absorption at 976 nm) of 50 cm is used as the gain medium pumped via a

980/1030 wavelength division multiplexer (WDM). A polarization independent isolator (ISO) guarantees the laser operating unidirectional in the ring cavity. A polarization controller (PC) was used to adjust optical polarization to obtain optimal operation state. The SA, SMF-GIMF-GIMF-SMF device was placed after the PC. A 99:1 coupler was inserted in the cavity as an output coupler. A 80:20 coupler was connected to a spectrometer and an oscilloscope respectively. The pulse train was measured by a 3 GHz photodetector connected to a 4 GHz oscilloscope (LeCroy Wave Runner 640Zi). The optical spectra were measured by an optical spectrum analyzer (OSA, ANDO AQ6317B). A 2Hz-2GHz RF spectrum analyzer (AV4021) is used to obtain the repetition rate of pulse train. The pulse width was obtained by APE pulseCheck 600 autocorrelator.

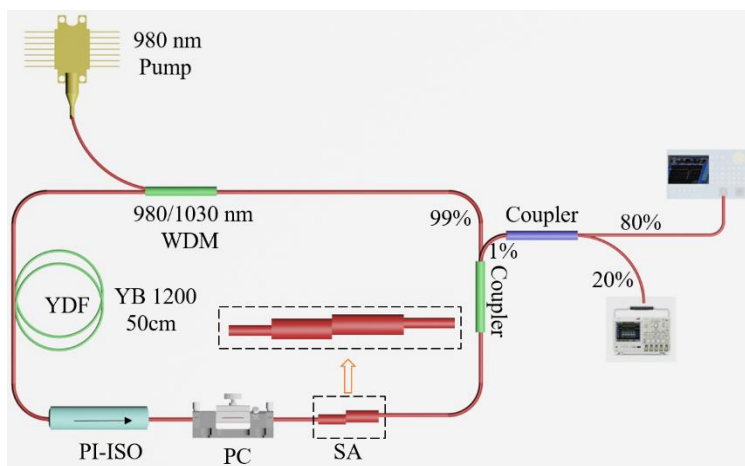

Fig. 3. Experimental setup of the fiber laser. LD: laser diode; WDM: wavelength division multiplexer; YDF: Yb-doped fiber; PI-ISO: polarization-independent isolator; PC: polarization controller; SA: SA; coupler: optical coupler.

The spectrum of the laser pulse is illustrated in Fig. 4 (a), which is centered at 1039 nm with steep spectral edges. Figure 4 (b) shows the mode-locked pulse train with a period of 32.67 ns measured by the oscilloscope. The inset of Fig. 4 (b) is oscilloscope trace with a time range of 2 μs. As shown in Fig. (c), the RF spectrum was measured by an RF spectrum analyzer. The RF peak is centered at 32.619 MHz corresponding to the resonant cavity length of the fiber laser about 3.066 m. The signal-to-noise ratio of pulses train as high as 65 dB indicates the high stability of the mode-locked state. The inset of Fig. 4 (c) is RF spectrum with frequency range from 0 to 1 GHz without any sidebands which also presents the stability of laser pulses. The mode-locked pulse was measured by the autocorrelator as shown in Fig. 4 (d). It displays the autocorrelation envelop of the laser pulse, showing a pulse width of 17.6 ps under Gaussian fitting, which corresponds to a peak power of 2.42 W. The time-bandwidth product (TBP) of the pulses is ~ 23.67, indicating that the output pulses are highly chirped.

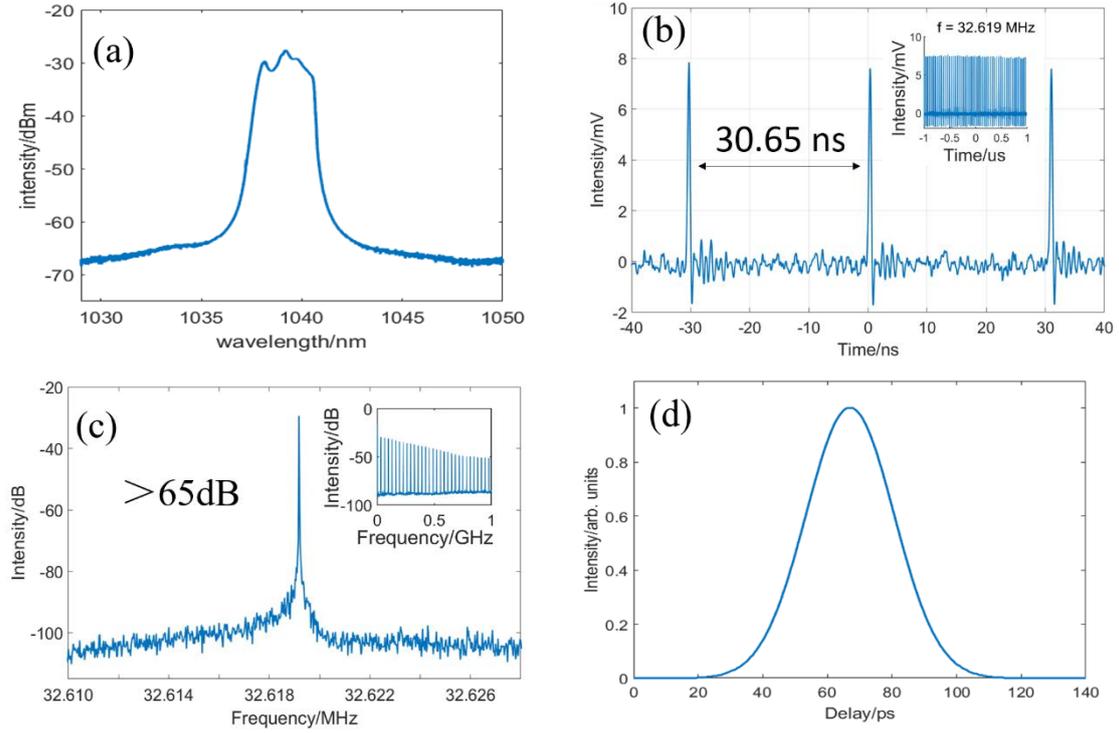

Fig. 4. Mode-locked pulse measurements. (a) Optical spectrum. (b) Oscilloscope trace of the single pulse emission (Inset: oscilloscope trace with a time duration of 2 μs ). (c) RF spectrum at the fundamental frequency of 32.619 MHz (Inset: RF spectrum over a frequency range of 0-1 GHz). (d) Autocorrelation trace

The mode-locked operation of the fiber laser is purely contributed by the SA structure. When the SA was removed, the laser could not generate any laser pulse train. Thus, it could be concluded that the mode-locked operating principle in our proposed system is not the nonlinear polarization rotation (NPR) effect.

The mode-locked state of fiber laser can be optimized through further study of the influence of the relative lengths of the 50 μm GIMF and 62.5 μm GIMF on the fiber laser threshold. As listed in Table I, the fiber laser threshold is dramatically fluctuated at different GIMF lengths. According to our experiments, the minimum fiber laser threshold of about 114 mW was achieved when the length of the 50 μm GIMF is 5 cm and the length of the 62.5 μm GIMF is 23 cm. Similarly, the output power of fiber laser varies dynamically with the GIMF lengths. The maximum average output power was achieved about 1.21 mW when the length of the 50 μm GIMF is 5 cm and the length of the 62.5 μm GIMF is 20 cm.

Table I. Properties of fiber laser threshold for the SMF-GIMF-GIMF-SMF structures with different lengths.

| Sample | Length (cm) 50μmGMIF | 62.5μm GMIF | Laser threshold (mW) | Output power (mW) |
|---|---|---|---|---|
| 1 | 5 | 17 | 162 | 0.98 |
| 2 | 5 | 20 | 143 | 1.21 |
| 3 | 5 | 21 | 132 | 1.02 |
| 4 | 5 | 23 | 114 | 1.10 |
| 5 | 5 | 29 | 174 | 1.14 |
| 6 | 3 | 20 | 189 | 1.06 |
| 7 | 9 | 20 | 136 | 1.00 |

4. **Summary**

In summary, we have developed a novel stable Yb-doped mode-locked fiber laser using the offset splicing technology between two kinds of graded index multimode fibers as the SA. The device has the advantages of low saturation intensity, easy fabrication, and low cost. The fiber laser generates ultrshort pulse trains with a pulse width of ∼17.6 ps, an average power of 1.1 mW at the repetition rate of 32.619 MHz. The fiber laser demonstrated has potential applications in ultrafast seed resources and nonlinear optics.